\documentclass[12pt,a4paper]{article}
\usepackage{amsmath}    %
\usepackage{amsfonts}   %
\usepackage{amssymb}    %
\usepackage{natbib}
\usepackage{rotate}
\usepackage{lscape}
\usepackage[flushleft]{threeparttable}
\usepackage{longtable}
\usepackage{colortbl}
\usepackage{supertabular}
\usepackage{mathrsfs}
\usepackage[brazil,english]{babel} 
\usepackage[latin1]{inputenc}   
\usepackage[T1]{fontenc}        
\usepackage{graphicx}
\DeclareGraphicsExtensions{.pdf,.gif,.jpg, ps, eps, png}

\usepackage{flafter}    
\usepackage{placeins}   

\usepackage[onehalfspacing]{setspace}   
\usepackage{geometry}
\usepackage{amsmath}
\usepackage{mathrsfs}
\usepackage{hyperref}

\newcommand{\bx}{\mathbf{x}}
\newcommand{\bz}{\mathbf{z}}

\newcommand{\bg}{\mathbf{g}} 
\newcommand{\bG}{\mathbf{G}} 
\newcommand{\bb}{\mathbf{b}} 
\newcommand{\bB}{\mathbf{B}}
\newcommand{\bkappa}{\mbox{\boldmath $\kappa$}}
\newcommand{\bbeta}{\mbox{\boldmath $\beta$}}

\newcommand{\bTheta}{\mbox{\boldmath $\Theta$}}

\newcommand{\bgamma}{\mbox{\boldmath $\gamma$}}
\newcommand{\bxi}{\mbox{\boldmath $\xi$}}
 
\newcommand{\bphi}{\mbox{\boldmath $\phi$}} 
\newcommand{\bpsi}{\mbox{\boldmath $\psi$}}

\usepackage{tablefootnote}

\begin{document}



\title{An Unified Semiparametric Approach to Model Lifetime Data with Crossing Survival Curves}
\author{Fábio Nogueira Demarqui\thanks{Departamento de Estatística, Universidade Federal de Minas Gerais, Brasil, email: fndemarqui@est.ufmg.br}  \\ 
Vinicius Diniz Mayrink\thanks{Departamento de Estatística, Universidade Federal de Minas Gerais, Brasil, email: vdm@est.ufmg.br} \\
Sujit Kumar Ghosh\thanks{Department of Statistics, North Carolina State University, USA, email: sujit.ghosh@ncsu.edu} }

\date{}

\maketitle

\begin{abstract}
The proportional hazards (PH), proportional odds (PO) and accelerated failure time (AFT) models have been widely used in different applications of survival analysis. Despite their popularity, these models are not suitable to handle lifetime data with crossing survival curves. In 2005, Yang and Prentice proposed a semiparametric two-sample strategy (YP model), including the PH and PO frameworks as particular cases, to deal with this type of data. Assuming a general regression setting, the present paper proposes an unified approach to fit the YP model by employing Bernstein polynomials to manage the baseline hazard and odds under both the frequentist and Bayesian frameworks. The use of the Bernstein polynomials has some advantages: it allows for uniform approximation of the baseline distribution, it leads to closed-form expressions for all baseline functions, it simplifies the inference procedure, and the presence of a continuous survival function allows a more accurate estimation of the crossing survival time. Extensive simulation studies are carried out to evaluate the behavior of the models. The analysis of a clinical trial data set, related to non-small-cell lung cancer, is also developed as an illustration. Our findings indicate that assuming the usual PH model, ignoring the existing crossing survival feature in the real data, is a serious mistake with implications for those patients in the initial stage of treatment. 
\end{abstract}


\section{Introduction}

\par Proportional hazards (PH) models \citep{Cox1972} have played a central role in many applications related to survival analysis. This class of models provides a flexible framework for time-to-event data and allows an easy interpretation of parameters, from the practical point of view. The main assumption of the PH models is the proportionality of the hazards ratios over time. When such assumption is not verified for the data, some alternatives are available, such as the proportional odds (PO) \citep{Bennett83} and the accelerated failure time (AFT) \citep{Kalb2002} models. However, none of these alternatives can accommodate survival data with crossing survival curves.

\par Crossing survival curves may arise due to several reasons. According to \cite{2013_Diao}, this can happen in clinical trials where a particular aggressive treatment (e.g. surgery) can have adverse effects at the beginning, but it may show beneficial results in the long run. Furthermore, as discussed in \cite{1984_Breslow}, crossing survival curves may also occur (see Figure \ref{fig_ht}) when a treatment has an early quick effect and it becomes similar or worse than the placebo treatment after a time period.  One example of this type of situation can be observed in the IRESSA Pan-Asia Study (IPASS). The IPASS clinical trial is a phase 3, open-label study, conducted with the primary objective of showing the noninferiority of the drug gefitinib compared to the option carboplatin-paclitaxelin as a first-line therapy for the intention-to-treat population patients with lung cancer \citep{2009_Mok}. 

The results of the IPASS clinical trial were firstly analyzed by \cite{2009_Mok}, who stated the superiority of gefitinib over carboplatin/paclitaxel. Gefitinib was approved in the U.S. in 2015 as a first-line therapy by the Food and  Drug Administration (FDA), and nowadays is marketed in over 90  countries. Nevertheless, the superiority of gefitinib over carboplatin/paclitaxel must be interpreted with caution, since the PH assumption is clearly violated for this data set, and any statistical analysis based on the PH  model, as in \cite{2009_Mok}, is not appropriate in this case. In addition, the inspection of the Kaplan-Meyer estimates of the survival curves provided in \cite{2009_Mok} indicates that, although gefitinib appears to have a superior performance over carboplatin/paclitaxel in the long run, the same does not seem to be true at earlier stages of treatment. In this fashion, a model capable of detecting treatment differences at initial and final periods of  follow up time would be desired. Furthermore, from the practical point of view, in order to increase patients' survival probabilities it is extremely important to take into account the time at which the treatments invert their effectiveness. 

Several approaches have been proposed in the literature to handle lifetime data with crossing survival curves. The most popular are the ones based on time-varying regression coefficients. However, this type of models has some drawbacks such as a larger number of parameters, implying in higher model complexity, and results harder to interpret. As an interesting alternative, \cite{2005_YangPrentice} proposed a semiparametric two-sample model (YP model hereafter) that includes the PH and PO frameworks as particular cases. In their model the baseline hazard function is not specified, and a pair of short and long-term hazards ratio parameters is included in the model to accommodate the feature of survival curves intersecting each other. A pseudo maximum likelihood approach is considered to estimate the parameters. Consistency and asymptotic normality of the estimators were evaluated in their work. \cite{2011_Yang} extended the estimation procedure in \cite{2005_YangPrentice} to separate and simultaneous inference on the hazards ratio function itself. The authors prove the consistency and asymptotic normality of estimates at a fixed time point. \cite{2012_Yang} proposed two omnibus tests to verify how appropriate is the YP model for the data. The first test is based on the martingale residuals and the second one accounts for the contrast between the nonparametric and model-based estimators for the survival function. \cite{2013_Diao} extended the two-sample YP model to a general regression setting with possibly time-dependent covariates. They have also developed an efficient likelihood-based estimation procedure and demonstrated the consistency, asymptotic normality and efficiency of the estimators. The YP model is also extended in \cite{2007_Tong} to accommodate current status survival data. \cite{2017_Zhang} considers the YP model to fit case II interval-censored data. 

In the past years, the Bernstein polynomials (BP) have gained visibility rising as an instrument for the statistical analyses in different applications. A thorough presentation of BP and its main mathematical aspects can be found in \cite{Lorentz12}. The use of BP for density estimation is a very common topic in the literature. This idea was proposed in \cite{Vitale75} and later explored in \cite{Petrone99}, \cite{Babu02} and \cite{Choud04}. Some authors have also considered the BP as an strategy to enhance regression modeling results; two interesting examples are \cite{Tenbusch97} and \cite{Chang07}. There are few studies using the BP to model survival data. A quick online search combining the keywords ``survival analysis" and ``Bernstein Polynomials" currently returns a short list of works. One of them is \cite{Chang05}, which considers the Bayesian framework to estimate the hazard function by assuming a Beta process prior and an homogeneous population due to the absence of covariates. In this reference, the polynomial degree is a random quantity that should be estimated. Another reference is \cite{2012_OsmanGhosh}, which is focused on the estimation of the hazard function in the context of crossing survival curves. The paper proposes a likelihood maximization procedure and considers the polynomial degree as a known quantity. A third reference is \cite{Chen14}, which investigates an accelerated hazard model incorporating the Dirichlet process and assuming a random polynomial degree. Their Bayesian analysis proposes a transformed Bernstein polynomial prior centered at standard parametric families (e.g. Weibull). More recently, \cite{Zhou17} presented an unified approach to handle arbitrarily censored spatial survival data in three semiparametric contexts: PH, PO and AFT. Baseline survival is modeled with the transformed Bernstein polynomial prior. This is essentially a Dirichlet process prior, assumed for the polynomials coefficients, having a parametric baseline distribution representing the center of the unknown survival function. \cite{QZhou17} considers semiparametric transformation models for interval-censored data. In their approach, Bernstein polynomials are applied to approximate the unknown baseline cumulative hazard function. 

The present paper takes advantage of the flexibility brought by the Bernstein polynomials as a representation for the hazard function or the odds function in the YP model. The BP modeling provides a continuous survival function and this feature allows a more accurate estimation of the time point where the curves are crossing. Note that, this accuracy is compromised when working with a stepwise survival curve, which is currently considered in \cite{2005_YangPrentice} and other references. A comprehensive simulation study is developed here to show that the BP modeling is in fact a good option for the analysis, being able to correctly capture the true shape of the survival function and estimate the regression coefficients, and the crossing survival times as well, with small bias. Our proposal configures a more appropriate model to handle the aforementioned IPASS data where the inversion of treatment performances is clear. In order to illustrate the gain in terms of analysis, we explore a reconstructed version of this data set provided in \cite{2015_Argyropoulos}.

Another interesting aspect investigated in this paper is related to the effects of more than one covariate. This has been previously mentioned in the literature, but never properly explored through simulation studies. According to \cite{2005_YangPrentice}, it is reasonable to assume in some applications that a subset of the covariates has the same coefficients for both short and long-term hazard ratios. In other words, the impact of those covariates is constant along the time and does not have a key contribution to explain the crossing behavior of the curves. As an example, for some diseases the gender of the patients may not be regarded as a factor modifying the decay speed of the survival curves, corresponding to treatment and control groups, between the short and long term periods. In this case, it would be more natural to consider the variable gender having the same effect on both hazards ratios. This assumption has important implications for the analysis, since it reduces the model complexity and the computational cost. The larger the subset of covariates with the same impact on both ratios, the more parsimonious is the model.

The main contributions of this paper are: \vspace{-5pt}
\begin{itemize}
 \item It proposes novel semiparametric approaches to model right-censored lifetime data with crossing survival curves. The main motivation is the YP model, which is improved here by using the BP to allow a continuous hazard or odds function.
 \item To the best of our knowledge, all semiparametric frameworks proposed in the literature to extend the YP model, and the original YP model itself, determine an increasing step function representation for the cumulative hazard or the odds function. Smoothness can be obtained when the BP are employed to model those baseline functions.
 \item The methodologies are investigated under the frequentist and Bayesian frameworks for a broaden presentation of models to be considered by all audiences in statistics.
 \item A comprehensive simulation study is developed. This involves more than one covariate to explain the long and short-term hazard ratios. The reader should note that only the single covariate scenario is investigated in important references such as \cite{2013_Diao} and \cite{2014_Barajas}.
 \item The Bernstein polynomials are used to model either the baseline hazard or the baseline odds of the YP model. All previous works are focused in only one of them; for example, \cite{2005_YangPrentice} and \cite{2014_Barajas} model the odds function and \cite{2013_Diao} considers the hazard function.
 \item The paper evaluates both strategies assuming all or only a subset of the covariates with distinct coefficients affecting the short and long-term hazard rations. The subset scenario sharing the same coefficients for these terms is only mentioned in \cite{2005_YangPrentice} and has never been explored through simulations.
\end{itemize}

This paper is organized as follows. The proposed versions of the YP models assuming the BP to represent the baseline hazard or the odds are described in Section \ref{sec_model}. This section also introduces notations and describes the scenarios with and without a subset of covariates having the same impact over the short and long-term ratios. A simulation study is conducted in Section \ref{sec_sim1} to evaluate the performance of the models under the two-sample scenario and the general regression setting. In Section \ref{sec_applic}, the analysis of the reconstructed IPASS data is developed to show the advantages of the proposed model. Finally, Section \ref{sec_conclusions} discusses the main conclusions and final remarks.

\section{Model Formulations} \label{sec_model}

This section presents the details related to the specification of four different versions of the YP model for crossing survival curves. The distinctions among the models are related to: ($i$) using the BP to handle the hazard function or the odds function and ($ii$) setting different short and long-term hazards ratios coefficients for all covariates or assuming a subset of covariates having the same coefficients for both ratios.     

\subsection{Yang and Prentice Model} \label{ssec_yp}

Let $T$ be a nonnegative random variable representing the time until the occurrence of an event of interest. In order to accommodate survival data with crossing survival curves, \cite{2005_YangPrentice} proposed the following model
\begin{eqnarray} \label{St_yp}
 S(t|\bTheta, \bz) = \left[ 1 + \frac{ \lambda }{ \theta }R_{0}(t)\right]^{-\theta},
\end{eqnarray}
where $\bz = (z_{1}, \ldots, z_{q})$ is a set of explanatory variables, $\lambda = \exp\{\bz \bpsi\}$ and $\theta = \exp\{\bz \bphi\}$, $\bpsi^\top = (\psi_{1}, \ldots, \psi_{q})$ and $\bphi^\top=(\phi_{1}, \ldots, \phi_{q})$ are vectors of regression coefficients, not including intercepts, $\bTheta=(\bpsi, \bphi)$, and $R_{0}(t)$ is a monotonic increasing function satisfying $\lim_{t \rightarrow 0}R_{0}(t) = 0$ and $\lim_{t \rightarrow \infty}R_{0}(t) = \infty$. 

The hazard function associated with \eqref{St_yp} takes the form
\begin{eqnarray} \label{ht_yp1}
 h(t|\bTheta, \bz) =  \frac{\lambda \theta \; R'_{0}(t)}{\lambda + \theta R_{0}(t)}, \;\;\; \mbox{where} \;\;\; R'_{0}(t) = \frac{d}{dt}R_{0}(t).
\end{eqnarray}
If $\bz = \mathbf{0}$, then \eqref{St_yp} reduces to 
\begin{eqnarray} \label{St0}
 S_{0}(t) = S(t|\bTheta, \mathbf{0}) = \frac{1}{1+R_{0}(t)},
\end{eqnarray}
which is referred to as the baseline survival function. Naturally, we can write $F_{0}(t) = 1 - S_{0}(t)$. In addition, this result and \eqref{St0} imply that $R_{0}(t) = F_{0}(t)/S_{0}(t)$, therefore, $R_{0}(t)$ corresponds to the baseline odds function. As a consequence, $R'_{0}(t) = h_{0}(t)/S_{0}(t)$,  where $h_{0}(t)$ = $-\frac{d}{dt}\ln S_{0}(t)$ is the baseline hazard function. 

The hazard function in \eqref{ht_yp1} can be rewritten in terms of the baseline functions $h_{0}(t)$, $F_{0}(t)$ and $S_{0}(t)$ as follows
\begin{eqnarray} \label{ht_yp2}
 h(t|\bTheta, \bz) =  \frac{\lambda\theta}{\lambda F_{0}(t)+\theta S_{0}(t)} \; h_{0}(t).
\end{eqnarray}
From expression \eqref{ht_yp2}, one can easily see that
\begin{eqnarray} \label{shortlong} 
 \lim_{t \rightarrow 0} \frac{h(t|\bTheta, \bz)}{h(t|\bTheta, \mathbf{0})} = \exp\{\bz\bpsi\} = \lambda, \nonumber \\
 \lim_{t \rightarrow \infty} \frac{h(t|\bTheta, \bz)}{h(t|\bTheta, \mathbf{0})} = \exp\{\bz\bphi\} = \theta. \nonumber 
\end{eqnarray}
Thus, the quantities $\lambda$ and $\theta$ can be interpreted as the short and long-term hazards ratios, respectively. In line with this notation, $\bpsi$ and $\bphi$ are regarded as the short and long-term regression coefficients. Another attractive feature of the YP model is the fact that it includes the PH (when $\bpsi = \bphi$) and PO (when $\bphi = \mathbf{0}$) models as particular cases. Moreover, it can be shown that crossing survival curves occur when $\psi_{j} \phi_{j} < 0$, i.e. when $\psi_{j}$ and $\phi_{j}$ have opposite signs for any $j = 1, \ldots, q$. 
\par According to \cite{2005_YangPrentice}, an alternative formulation of their model can be obtained by assuming that the effects of some covariates do not change through time. Although appealing from the practical point of view in many real situations, as far as we known, such alternative formulation has not been addressed in the literature until now. The alternative formulation of the YP model can be obtained as follows. Denote by $\bz^* = (\bz, \bx)$ a $1 \times q^{*}$ vector of covariates, with $q^{*}=q+p$, where $\bz$ is a $1 \times q$ vector incorporating those explanatory variables whose effects are expected to change between the short and long-term periods, and $\bx$ is a $1 \times p$ vector including those covariates whose effects are expected to remain constant over time. Let $\bpsi^* = (\bpsi, \bbeta)$ and $\bphi^{*}=(\bphi, \bbeta)$, where $\bpsi$ and $\bphi$ are $q \times 1$ vectors containing the short and long-term regressions coefficients associated with $\bz$, respectively, and $\bbeta$ is a $p \times 1$ vector of constant-time regression coefficients associated with $\bx$. Then, replacing $\bTheta$ and $\bz$ by $\bTheta^{*}=(\bpsi^{*}, \bphi^{*}) \equiv (\bpsi, \bphi, \bbeta)$ and $\bz^{*}$ in the expressions presented above, it is straightforward to show that the new hazard and survival functions associated with the alternative formulation of the YP model are expressed, respectively, as
\begin{eqnarray}
 h(t|\bTheta^{*}, \bz^{*}) = h(t|\bTheta, \bz)e^{\bx\bbeta},  \label{yp_alt} \\
 S(t|\bTheta^{*}, \bz^{*}) = S(t|\bTheta, \bz)^{e^{\bx\bbeta}}; \nonumber
\end{eqnarray}
where $h(t|\bTheta, \bz)$ and $S(t|\bTheta, \bz)$ were previously defined; see the expressions in (\ref{ht_yp1}), (\ref{ht_yp2}) and (\ref{St_yp}). Thus, given $\bz$, the proportional hazards assumption holds for \eqref{yp_alt}.

\par In order to fit the YP model, one can either focus on modeling the baseline hazard or the baseline odds function. This task has been traditionally done by means of step functions with jumps defined by the observed failure times \citep[see, for instance,][]{2005_YangPrentice,2007_Tong,2013_Diao,2014_Barajas}. 
One drawback of these approaches is the fact that the resulting survival functions are also step functions, with jumps defined by the observed failure times. From the practical point of view, a continuous survival function would be convenient, since it allows a more accurate identification of the time at which the survival curves intersect each other. As it will be demonstrated ahead, this issue can be easily addressed by using Bernstein polynomials, which are briefly described next. 

\subsection{Bernstein Polynomials in Survival Analysis} \label{ssec_bp}

Let $C(\cdot)$ be a continuous function on the interval $(0, \tau]$. The Bernstein polynomial of  order $m$ evaluated in $t \in (0,\tau]$, with base $\bB_{m} = (B_{0,m}, B_{1,m}, \ldots, B_{m,m})$, and coefficients $\bb_{m} = (b_{0,m}, b_{1,m}, \ldots, b_{m,m})$, intended to approximate a given function $C(\cdot)$, is defined by
\begin{eqnarray}
  B_{m}(t; C) = \sum_{k=0}^{m} b_{k,m} \; B_{k,m}(t),
\end{eqnarray}
where $B_{k,m} = \binom{m}{k} (t/\tau)^{k} (1-t/\tau)^{m-k}$ and $b_{k,m} = C(k \tau/m)$, for $k = 0, 1, \ldots, m$. The literature related to the BP shows that $B_{m}(\cdot ; C) \rightarrow C(\cdot)$ uniformly on $(0, \tau]$ as $m \rightarrow \infty$. Details can be found in \cite{Lorentz12} and references therein.

The first derivative of $B_{m}(t;C)$, denoted by $b_{m}(t; C)$, can be written as 
\begin{eqnarray} \label{bern_deriv}
 \hspace{-0.5cm} & b_{m}(t ; C) =  \sum_{k=1}^{m} \left[C\left(\frac{k}{m} \tau \right) - C\left(\frac{k-1}{m} \tau \right)\right] \; \frac{f_{B}(t/\tau|k, m-k+1)}{\tau}, 
\end{eqnarray}
where $f_{B}$ is the density of the beta distribution evaluated at $t/\tau$ and having shape parameters $k$ and $m-k+1$. The literature also shows that $b_{m}(\cdot; C) \rightarrow c(\cdot)$ as $m \rightarrow \infty$, uniformly on $(0, \tau]$, where $c(t) = dC(t)/dt$.

According to \cite{2012_OsmanGhosh} the hazard function can be modeled as follows
\begin{eqnarray} \label{ht_bp}
 h(t|\bgamma) \; = \; \sum_{k=1}^{m} \gamma_{k} \; g_{k,m}(t) \; = \; \bgamma \; \bg_{m}(t), \;\; \mbox{for} \; t \geq 0;
\end{eqnarray}
where $\bgamma = (\gamma_{1}, \ldots, \gamma_{m})$ is the vector of unknown coefficients satisfying $\gamma_{k} \geq 0$, for $k = 1, \ldots m$. In addition, $\bg_{m}(t) = (g_{1,m}(t), \ldots, g_{m,m}(t))^\top$ is the vector of baseline functions evaluated at the time point $t$. In this case, $g_{k,m}(t) \geq 0$ and $\int_0^\infty g_{k,m}(u) du < \infty$ for all $k \leq m$. This result is connected with the one expressed in \eqref{bern_deriv}, therefore, we can identify $g_{k,m}(t) = f_{B}(t/\tau| k, m-k+1)/\tau$ and $\gamma_{k} = C\left( \frac{k}{m}\tau \right)-C\left( \frac{k-1}{m}\tau \right)$.

The cumulative hazard function is given by 
\begin{eqnarray} \label{Ht_bp}
 H(t|\bgamma) & = & \int_{0}^{t} \sum_{k=1}^{m} \gamma_{k} \; g_{k,m}(u) du \; = \; \sum_{k=1}^{m} \gamma_{k} \int_{0}^t g_{k,m}(u) du \nonumber \\
 & = & \sum_{k=1}^{m} \gamma_{k} \; G_{k,m}(t) \; = \; \bgamma \bG_{m}(t), \;\; \mbox{for} \; t \geq 0.
\end{eqnarray}
Here, $G_{k,m}(t) = F_{B}(t/\tau| k, m-k+1)$, for $k = 1, \ldots, m$, and $F_{B}$ represents the cumulative distribution function (c.d.f.) of the Beta($k$, $m-k+1$) evaluated at $t/\tau$. The monotonicity of $H(t|\bgamma)$ is ensured by the restrictions: $\gamma_{k} \geq 0$ and $g_{k,m}(t) \geq 0$, for $t > 0$ and $k = 1, \ldots, m$.

Formally, for the context of survival analysis, one must set $\tau < \infty$ such that $\tau = \inf\{t : S(t) = 0\}$. However, the specification in \eqref{Ht_bp} does not satisfy $H(\tau|\bgamma) = \infty$. It is necessary to apply a tail adjustment to correct this issue. \cite{2012_OsmanGhosh} suggest choosing $\hat{\tau} = \max\{ t_1, \ldots, t_n \}$, where $t_i$ is the $i$-th observed time point, and then assuming a constant hazard function for $t > \hat{\tau}$, since there is no data information about the failure time distribution in this region. The hazard function can be written as follows
\begin{eqnarray*}
 h^*(t|\bgamma) = \left\{
 \begin{array}{l}
   h(t|\bgamma), \;\; \mbox{if} \;\; 0 \leq t < \hat{\tau}, \\
   m \; \gamma_{m}/\hat{\tau}, \;\; \mbox{if} \;\; t \geq \hat{\tau}.
 \end{array}
 \right.
\end{eqnarray*}
The corresponding cumulative hazard function is
\begin{eqnarray*}\displaystyle
  H^*(t|\bgamma) = \left\{
  \begin{array}{l}
    H(t|\bgamma), \;\; \mbox{if} \;\; 0 \leq t < \hat{\tau}, \\
    H(t|\bgamma) + m\;(t-\hat{\tau}) \; \gamma_{m}/\hat{\tau}, \;\; \mbox{if} \;\; t \geq \hat{\tau}.
  \end{array}
  \right.
\end{eqnarray*}
It can be shown that $\int_0^\infty h^{*}(t|\bgamma_m, m) \; dt = \infty$, therefore, the mentioned tail adjustment determines a valid hazard function. The indicated $\hat{\tau}$ ensures that all observations impacting the likelihood are handled by the BP, thus the properties of the survival model considered for the application are unaffected. Note that, as the sample size increases ($n \rightarrow \infty$) the value $\hat{\tau} \rightarrow \tau = \inf\{ t > 0: S(t) = 0 \}$ in probability.

Since $R_0(t)$ behaves like $H_0(t)$, it is reasonable to consider the same structure adopted for $H_{0}(t)$ to model the baseline odds $R_0(t)$. In line with this idea, we set
\begin{eqnarray} \label{Rt_bp}
  R_0'(t|\bxi) = \sum_{k=1}^{m} \xi_{k} \; g_{k,m}(t) \;\; \mbox{and} \;\; R_0(t|\bxi) = \sum_{k=1}^{m} \xi_{k} \; G_{k,m}(t),
\end{eqnarray}
where $\bxi_m = (\xi_{1,m}, \ldots, \xi_{m,m})$ is the vector of unknown coefficients, with $\xi_{k,m} \geq 0$ for $k=1, \ldots, m$. The notation $\bxi_m$ is introduced to make distinction with respect to $\bgamma_m$ specified for the hazard function case. The definitions of $g_{k,m}(t)$ and $G_{k,m}(t)$ did not change here. 

\subsection{Proposed Models} \label{ssec_models}

We propose to model both $h_{0}(\cdot)$ and $R_{0}(\cdot)$ using the Bernstein polynomials. Two versions are considered for the YP model. The first one is the original formulation presented in the literature, which assumes different effects over the short and long-term hazards ratios for all covariates;  hereafter this version is referred to as ``original YP model''. The second option assumes that the effect over the short and long-term ratios is the same for a subset of the covariates; this version will be denoted as ``alternative YP model''. The four configurations of models to be explored are indicated in Table \ref{tab_models}.

\begin{table}[!h]
\centering \footnotesize
\caption{Model specifications being explored in the present paper.} \label{tab_models} \vspace{5pt}
\begin{tabular}{lclc} \hline
 YP version  & Model & Parameters &BP for \\  \hline
 original    & $M_{1}$ & $\bTheta_{1}=(\bpsi, \bphi, \bgamma)$ & $h_0(\cdot)$ \\
             & $M_{2}$ & $\bTheta_{2}=(\bpsi, \bphi, \bxi)$ &  $R_0(\cdot)$ \\ \hline
 alternative & $M_{1}^{*}$ & $\bTheta_{1}^{*}=(\bpsi, \bphi, \bbeta, \bgamma)$ &  $h_0(\cdot)$ \\
             & $M_{2}^{*}$ & $\bTheta_{2}^{*}=(\bpsi, \bphi, \bbeta, \bxi)$ &  $R_0(\cdot)$ \\  \hline \\
\end{tabular}
\end{table}

In $M_1$ and $M_{1}^{*}$, the hazard function takes the form in \eqref{ht_yp2} and the BP representation in \eqref{ht_bp} is assumed for $h_0(t)$. As a consequence, the baseline cumulative hazard $H_0(t)$ is given in \eqref{Ht_bp}. Regarding the baseline survival function, recall that $S_0(t) = \exp\{-H_0(t)\}$. The models $M_2$ and $M_{2}^{*}$ use the second formulation of the hazard function presented in \eqref{ht_yp1}. Here, the BP representations in \eqref{Rt_bp} are considered for the baseline odds function and its derivative.

\par Consider a random sample of size $n$ of independent elements and denote by $T_{i}$ and $C_{i}$, respectively, the failure and censoring times. Recall the previous notation and let $\bz_{i}$ and $\bx_{i}$, for $i = 1, \cdots, n$, be two vectors of explanatory variables. Assume that the censoring mechanism is non-informative. In addition, assume the failure times are right-censored so that $Y_{i} = \min\{T_{i}, C_{i}\}$ establishes the observed time for subject $i$. Let $\delta_{i} = I_{\{T_{i} \leq C_{i}\}}$ be the failure indicator function. Further denote $D^{*}=\{(y_{i}, \delta_{i}, \bz_{i}, \bx_{i}), i=1, \cdots, n\}$ as the full set of observed data. Then, under the alternative formulation of the YP model, a general expression for the likelihood function can be written as
\begin{align} \label{lik1}
  & L(\bTheta_{r}^{*}|D^{*}) = & \nonumber \\
  & \prod_{i=1}^{n} \left[h(t_i |\bTheta_{r}, \bz_{i})e^{\bx_{i}\bbeta}\right]^{\delta_i} \left[ 1 + \frac{ \lambda_i }{ \theta_i }R_{0}(t_i|\bkappa_{r})\right]^{-\theta_ie^{\bx_{i}\bbeta}},
\end{align}
where $\bTheta_{r}^{*}=(\bpsi, \bphi, \bbeta, \bkappa_{r})$, $r = 1$ or $2$, denotes the set of parameters to be estimated. We set $\bkappa_{1} = \bgamma$ or $\bkappa_{2} = \bxi$, depending on choice of the baseline (hazard or odds) function to be modeled by the Bernstein polynomials. Specifically, the likelihood function associated with model $M_{1}^{*}$ is obtained by setting: $h_{0}(t_{i})=h_{0}(t_{i}|\bgamma) = \sum_{k=1}^{m}\gamma_{k,m}g_{k,m}(t)$, $H_{0}(t_{i})=H_{0}(t_{i}|\bgamma) = \sum_{k=1}^{m} \gamma_{k,m}G_{k,m}(t)$, $S_{0}(t_{i}|\bgamma) = \exp\{-H_{0}(t_{i}|\bgamma)\}$, $F_{0}(t_{i}|\bgamma) = 1-S_{0}(t_{i}|\bgamma)$ and $R_{0}(t_{i}|\bgamma) = F_{0}(t_{i}|\bgamma)/S_{0}(t_{i}|\bgamma)$. For Model $M_{2}^{*}$, the likelihood function is configured with: $R'_{0}(t_{i})=R'_{0}(t_{i}|\bxi) = \sum_{k=1}^{m}\xi_{k,m}g_{k,m}(t)$ and $R_{0}(t_{i}) = R_{0}(t_{i}|\bxi)=\sum_{k=1}^{m} \xi_{k,m}G_{k,m}(t)$.
\par If no covariates are expected to have their effects constant over time, then $\bx_{i} = \mathbf{0}$ and expression \eqref{lik1} reduces to
\begin{eqnarray} \label{lik2}
  L(\bTheta_{r}|D) = \prod_{i=1}^{n} h(t_i |\bTheta_{r}, \bz_i)^{\delta_i} \left[ 1 + \frac{ \lambda_i }{ \theta_i }R_{0}(t_i|\bkappa_{r})\right]^{-\theta_i},
\end{eqnarray}
where $D=\{(y_{i}, \delta_{i}, \bz_{i}), i=1, \cdots, n\}$, $\bTheta_{r}=(\bpsi, \bphi, \bkappa_{r})$ with $r = 1$ or $2$. In this case, we have the general expression for the likelihood function associated with the original formulation of the YP model.
\par The closed form for the likelihood functions given in \eqref{lik1} and \eqref{lik2} allows us to easily employ likelihood-based methods to estimate parameters and related quantities. In order to determine the maximum likelihood (ML) estimates, one can apply direct maximization of the log-likelihood function $l(\bTheta) = \log L\left(\bTheta;D\right)$ by using the quasi-Newton BFGS method available in standard statistical softwares such as \texttt{R} \citep{cran} and SAS (\url{www.sas.com}).

It is worth noting that, although the observed Fisher information matrix, provided by the BFGS method, can be used to obtain point and interval estimates, it is not straightforward to determine an interval estimate for the crossing survival time, denoted here as $t^{*}$. The main reason for this is the fact that it does not exist a closed form expression for the standard error of the estimator $\hat{t}^{*}$. A possible solution is to numerically solve the non-linear equation $S_{0}(t^{*}) - S_{1}(t^{*}) = 0$. In the present paper, we circumvent this issue by proposing a non-parametric bootstrap method allowing to infer the quantities of interest.

Under the Bayesian framework, prior distributions expressing our initial uncertainty about the unknown quantities must be specified. We assume that $\bgamma$ (or $\bxi$), $\bpsi$, $\bphi$ (and $\bbeta$ for the alternative formulation of the YP model) are all independent such that: $\log(\gamma_{k}) \sim N(\mu_{\gamma}, \sigma_{\gamma})$ and $\log(\xi_{k}) \sim N(\mu_{\xi}, \sigma_{\xi})$, for $k = 1, \cdots, m$, $\psi_{j} ~ \sim N(\mu_{\psi}, \sigma_{\psi})$, $\phi_{j} ~ \sim N(\mu_{\phi}, \sigma_{\phi})$, for $j = 1, \cdots, q$ and, finally, $\beta_{l} ~ \sim N(\mu_{\beta}, \sigma_{\beta})$ for $l = 1, \cdots, p$. The assumption of independence is made here to guarantee a fair comparison with the results provided by those models fitted under the ML framework. The parameters in $\bgamma$ and $\bxi$ have a positive support, but we choose to model them in the log scale due to computational reasons. We emphasize, however, that other prior distributions for $\bgamma$ and $\bxi$, inducing some sort of dependency among their components, for example, can be specified. This is beyond the scope of the present paper, thus it is left for future work.

\section{Simulation Study} \label{sec_sim1}
%
\par In this section, we present a comprehensive Monte Carlo (MC) simulation study to evaluate the performance of the proposed models in terms of estimation of the regression coefficients and the crossing survival time point. The analysis is divided in two scenarios: 
\begin{itemize}
 \item Scenario I: a two-sample problem, with focus on the estimation of the regression coefficients and the crossing survival time;
 \item Scenario II: a general regression setting involving four covariates is investigated, with focus on the estimation of the regression coefficients. In this case, the original and the alternative formulations of the YP model are considered.
\end{itemize}

\par For each scenario, $1000$ MC replications of data sets, with sample size $n = 500$, were generated by assuming a Weibull baseline survival function $S_{0}(t|\alpha,\gamma)=\exp\left\{-\gamma t^{\alpha}\right\}$, with $\alpha=1.5$ and $\gamma=0.05$. Censoring times were originated from an uniform distribution $U(0,\nu)$, with $\nu$ chosen in a way that the censoring rate corresponds to approximately 30\% of the observed data. Under the Bayesian framework, for each MC replica, posterior samples of size $4000$ for the quantities of interest were drawn, using the NUTS algorithm \citep{2014_NUTS}, after running $4$ chains of length $2000$, with the first $1000$ iteration being discarded as a warm-up period. All prior distributions were set to have mean 0 and standard deviation 4. For comparison purposes, for each MC replica, bootstrap estimates for some target elements were obtained based on $4000$ bootstrap samples. All models were fitted by calling \texttt{Stan} \citep{Stan} from \texttt{R} \citep{cran} using the package \texttt{rstan} \citep{rstan}.

\par In order to generate the data sets for Scenario I, the following short-term and long-term linear predictors were used:
\begin{eqnarray*}
\log(\lambda_{i}) = +2.0z_{i} \quad \mbox{and} \quad \log(\theta_{i})  = -1.0z_{i},
\end{eqnarray*}
where $z_{i} \sim Bern(0.5)$. Note that the short and long-term regression coefficients have opposite signs, implying that the survival curves will cross. This fact allows us to verify the performance of the proposed models in the estimation of crossing survival time point. 

\begin{table}
\centering \footnotesize
\caption{Monte Carlo simulation study for Scenario I (two-sample). Estimate (est.), average standard error (se.), standard deviation of the estimates (sde.), relative bias (rb in \%) and coverage probability (nominal level 95\%) are presented for the coefficients and the crossing survival time ($t^*$).}
\begin{tabular}{rrrrrrrrrrrrrr} \hline 
   $M_1$ & & & \multicolumn{5}{c}{ML approach} & & \multicolumn{5}{c}{Bayesian approach} \\ \cline{4-8} \cline{10-14} 
       & true & & est. & se. & sde. & rb & cov. &  & est. & se. & sde. & rb & cov. \\   \hline 
$\psi$ & 2.00 & & 2.12 & 0.31 & 0.30 & 6.01 & 93.90 &  & 2.07 & 0.29 & 0.29 & 3.53 & 94.70 \\ 
$\phi$ & -1.00 & & -1.04 & 0.15 & 0.14 & -3.75 & 94.60 &  & -1.03 & 0.14 & 0.14 & -2.94 & 94.30 \\ 
$ t^{*}$ & 47.90 & & 49.22 & 9.81 & 11.04 & 2.76 & 96.0 &  & 48.49 & 11.20 & 10.14 & 1.23 & 94.0 \\ 
   \hline
  $M_2$ & & & \multicolumn{5}{c}{ML approach} & & \multicolumn{5}{c}{Bayesian approach} \\ \cline{4-8} \cline{10-14} 
       & true & & est. & se. & sde. & rb & cov. &  & est. & se. & sde. & rb & cov. \\   \hline 
   \hline   
$\psi$ & 2.00 & & 2.11 & 0.31 & 0.30 & 5.56 & 94.30 &  & 2.04 & 0.29 & 0.29 & 1.97 & 94.40 \\ 
$\phi$ & -1.00 & & -1.03 & 0.15 & 0.14 & -3.30 & 94.40 &  & -1.01 & 0.14 & 0.14 & -0.86 & 95.70 \\ 
$t^{*}$   & 47.90 & & 49.39 & 9.91 & 11.18 & 3.11 & 96.0 &  & 49.57 & 11.62 & 10.16 & 3.48 & 96.0 \\  \hline \\  
\end{tabular}
\label{tab_scen1}
\end{table}

\begin{table}
\centering \footnotesize
\caption{Carlo simulation study for Scenario II (general regression). Estimate (est.), average standard error (se.), standard deviation of the estimates (sde.), relative bias (rb in \%) and coverage probability (nominal level 95\%) are presented for the coefficients.}
\begin{tabular}{rrrrrrrrrrrrrr} \hline 
  $M_{1}$ & & & \multicolumn{5}{c}{ML approach} & & \multicolumn{5}{c}{Bayesian approach} \\ \cline{4-8} \cline{10-14} 
       & true & & est. & se. & sde. & rb & cov. &  & est. & se. & sde. & rb & cov. \\   \hline 
  $\psi_{1}$ &  2.00 & & 2.10 & 0.23 & 0.22 & 4.83 & 93.10 &  & 2.08 & 0.21 & 0.23 & 4.11 & 91.60 \\ 
  $\psi_{2}$ & -0.50 & & -0.53 & 0.10 & 0.09 & -5.34 & 93.70 &  & -0.53 & 0.10 & 0.10 & -6.08 & 93.10 \\ 
  $\psi_{3}$ & 1.50 & & 1.56 & 0.18 & 0.17 & 4.01 & 95.20 &  & 1.54 & 0.17 & 0.17 & 2.86 & 94.70 \\ 
  $\psi_{4}$ & -1.50 & & -1.55 & 0.11 & 0.10 & -3.62 & 92.90 &  & -1.54 & 0.11 & 0.11 & -2.50 & 94.00 \\ 
  $\phi_{1}$ & -1.00 & & -1.02 & 0.20 & 0.20 & -1.69 & 94.10 &  & -1.03 & 0.20 & 0.22 & -2.81 & 93.70 \\ 
  $\phi_{2}$ & 1.00 & & 1.04 & 0.15 & 0.14 & 4.39 & 94.00 &  & 1.03 & 0.14 & 0.16 & 2.98 & 94.80 \\ 
  $\phi_{3}$ & 1.50 & & 1.59 & 0.31 & 0.28 & 5.88 & 96.40 &  & 1.52 & 0.29 & 0.29 & 1.31 & 95.50 \\ 
  $\phi_{4}$ & -1.50 & & -1.58 & 0.22 & 0.20 & -5.26 & 93.70 &  & -1.54 & 0.21 & 0.24 & -2.57 & 95.10 \\ 
   \hline
  $M_{2}$ & & & \multicolumn{5}{c}{ML approach} & & \multicolumn{5}{c}{Bayesian approach} \\ \cline{4-8} \cline{10-14} 
       & true & & est. & se. & sde. & rb & cov. &  & est. & se. & sde. & rb & cov.  \\   \hline 
   \hline   
  $\psi_{1}$ & 2.00 & & 2.09 & 0.23 & 0.22 & 4.72 & 92.50 &  & 2.05 & 0.21 & 0.22 & 2.40 & 93.00 \\ 
  $\psi_{2}$ & -0.50 & & -0.53 & 0.10 & 0.09 & -5.24 & 93.60 &  & -0.51 & 0.09 & 0.09 & -1.73 & 94.40 \\ 
  $\psi_{3}$ & 1.50 & & 1.56 & 0.18 & 0.17 & 3.97 & 94.90 &  & 1.54 & 0.17 & 0.17 & 3.03 & 95.50 \\ 
  $\psi_{4}$ & -1.50 & & -1.56 & 0.11 & 0.10 & -3.73 & 92.70 &  & -1.55 & 0.10 & 0.10 & -3.11 & 93.90 \\ 
  $\phi_{1}$ & -1.00 & & -1.01 & 0.20 & 0.20 & -0.84 & 94.10 &  & -0.96 & 0.19 & 0.19 & 4.38 & 94.30 \\ 
  $\phi_{2}$ & 1.00 & & 1.04 & 0.15 & 0.14 & 3.64 & 94.20 &  & 1.01 & 0.14 & 0.14 & 0.73 & 96.00 \\ 
  $\phi_{3}$ & 1.50 & & 1.57 & 0.31 & 0.28 & 4.89 & 96.20 &  & 1.50 & 0.29 & 0.26 & 0.12 & 96.30 \\ 
  $\phi_{4}$ & -1.50 & & -1.56 & 0.21 & 0.20 & -3.74 & 95.30 &  & -1.40 & 0.19 & 0.18 & 6.80 & 90.70 \\    
   \hline 
  $M_{1}^{*}$ & & & \multicolumn{5}{c}{ML approach} & & \multicolumn{5}{c}{Bayesian approach} \\ \cline{4-8} \cline{10-14} 
       & true & & est. & se. & sde. & rb & cov. &  & est. & se. & sde. & rb & cov. \\   \hline 
  $\psi_{1}$ & 2.00 & & 2.08 & 0.20 & 0.21 & 4.11 & 91.90 &  & 2.07 & 0.19 & 0.21 & 3.40 & 92.20 \\ 
  $\psi_{2}$ & -0.50 & & -0.52 & 0.09 & 0.09 & -4.34 & 93.30 &  & -0.53 & 0.09 & 0.09 & -5.07 & 92.80 \\ 
  $\phi_{1}$ & -1.00 & & -1.01 & 0.19 & 0.18 & -1.43 & 94.00 &  & -1.03 & 0.18 & 0.18 & -2.67 & 93.50 \\ 
  $\phi_{2}$ & 1.00 & & 1.03 & 0.13 & 0.13 & 3.31 & 94.40 &  & 1.03 & 0.13 & 0.13 & 2.63 & 94.60 \\ 
  $\beta_{1}$ & 1.50 & & 1.55 & 0.12 & 0.12 & 3.27 & 93.40 &  & 1.52 & 0.12 & 0.12 & 1.56 & 94.90 \\ 
  $\beta_{2}$ & -1.50 & & -1.55 & 0.08 & 0.08 & -3.19 & 90.20 &  & -1.53 & 0.08 & 0.08 & -1.92 & 94.50 \\
   \hline   
  $M_{2}^{*}$ & & & \multicolumn{5}{c}{ML approach} & & \multicolumn{5}{c}{Bayesian approach} \\ \cline{4-8} \cline{10-14} 
       & true & & est. & se. & sde. & rb & cov. &  & est. & se. & sde. & rb & cov. \\   \hline 
  $\psi_{1}$ &  2.00 & & 2.08 & 0.20 & 0.21 & 3.79 & 92.30 &   & 2.00 & 0.19 & 0.20 & 0.08 & 93.20 \\ 
  $\psi_{2}$ & -0.50 & & -0.52 & 0.09 & 0.09 & -4.08 & 93.30 &  & -0.50 & 0.09 & 0.09 & 0.66 & 94.50 \\ 
  $\phi_{1}$ & -1.00 & & -1.01 & 0.18 & 0.18 & -0.70 & 94.10 &  & -0.93 & 0.18 & 0.18 & 6.66 & 93.00 \\ 
  $\phi_{2}$ & 1.00 & & 1.03 & 0.13 & 0.13 & 3.06 & 94.20 &   & 1.02 & 0.13 & 0.13 & 2.28 & 95.40 \\ 
  $\beta_{1}$ & 1.50 & & 1.55 & 0.12 & 0.12 & 3.08 & 94.00 &   & 1.53 & 0.12 & 0.12 & 2.01 & 95.30 \\ 
  $\beta_{2}$ & -1.50 & & -1.54 & 0.08 & 0.08 & -2.96 & 91.00 &  & -1.50 & 0.08 & 0.07 & 0.24 & 95.80 \\ 
   \hline \\   
\end{tabular}
\label{tab_scen2}
\end{table}

In the investigations conducted in this section for artificial data, we choose to explore a measurement accounting for the distance between the reported estimate and the true value of the parameter. This is called relative bias and it has the following formulation:
$$
 \text{rb}(\kappa) = 100 \ (\hat{\kappa} - \kappa_{\tiny \mbox{true}}) \ / \ |\kappa_{\tiny \mbox{true}}|.
$$
In this case consider: $\kappa$ is a generic parameter, $\hat{\kappa}$ is the maximum likelihood or posterior estimate and $\kappa_{\tiny \mbox{true}}$ is the target true value. The relative bias can be seen as a ratio between the estimation error and the magnitude of the true value. Negative and positive results indicate underestimation and overestimation, respectively. The fraction is multiplied by $100$ to adjust scale leading to a quantity indicating a percentage representing how big is the error with respect to the magnitude of the true value. This quantity is often used in the survival analysis literature.

The variability or uncertainty related to the point estimates are expressed, in this analysis based on MC replications, through the average standard error (se) and the standard deviation of the estimates (sde). For a given parameter $\kappa$, the average standard error has the formulation $\sum_{j=1}^{1000} \text{se}_j(\kappa) / 1000$, where $\text{se}_j(\kappa)$ is the standard error (or posterior standard deviation) obtained for $\kappa$ in the $j$-th sample of the MC scheme. On the other hand, the sde is given by $\sum_{j=1}^{1000} (\hat{\kappa}_j - \bar{\kappa})^2/999$, with $\hat{\kappa}_j$ being the estimated $\kappa$ in the $j$-th sample and $\bar{\kappa} = \sum_{j=1}^{1000} \hat{\kappa}_j / 1000$.

Table \ref{tab_scen1} presents the MC simulation results for Scenario I. As it can be seen, models $M_{1}$ and $M_{2}$ have a similar performance in terms of estimation. The results also suggest that the Bayesian framework provides slightly better estimates than the ML case. Overall, Table \ref{tab_scen1} shows small relative biases (absolute maximum 6.01\%) and coverage probabilities close to the nominal level of $95\%$. The coefficient $\phi$ tend to be underestimated (rb $< 0$) and the remaining quantities are overestimated (rb $> 0$).

The data generating procedure, associated with Scenario II, considers the short and long-term linear predictors as follows:
\begin{equation}
\begin{array}{l} 
\log(\lambda_{i}) = +2.0z_{1i} - 0.5z_{2i} + 1.5z_{3i} - 1.5z_{4i},  \\ 
\log(\theta_{i})  = -1.0z_{1i} + 1.0z_{2i} + 1.5z_{3i} - 1.5z_{4i}; 
\end{array} \nonumber
\end{equation}
where $z_{1i} \sim Bern(0.5)$, $z_{2i} \sim N(0,1)$, $z_{3i} \sim Bern(0.5)$ and $z_{4i} \sim N(0, 1)$, for $i = 1, \cdots, n$. It is important to highlight the fact that we choose opposite signs for the first two short and long-term coefficients, which implies that the survival curves will have an intersection at some intermediate point within the period where the data is generated. In addition, note that all coefficients are not time dependent. This configuration is chosen to enable a fair comparison among models based on the original and the alternative formulations of the YP model previously discussed. 

\par The conclusions from the analysis of Scenario II are similar to those obtained from Scenario I. According to Table \ref{tab_scen2}, all models being examined behave well, regardless of the approach considered for estimation. Note that again slightly better estimates can be observed for the models fitted under the Bayesian framework. Furthermore, the models $M_{1}^*$ and $M_{2}^*$ (alternative formulation) outperform the options $M_{1}$ and $M_{2}$ (original formulation) in terms of relative bias. This result is in fact expected, due to the procedure adopted to generate the data. 
In summary, the models $M_{1}$ and $M_{2}$ show robustness for estimation, despite their greater complexity.

\par When moving from Table \ref{tab_scen1} to \ref{tab_scen2}, it is possible to detect a small decrease in the overall performance of the models in terms of relative biases and coverage probability. This result is expected and can be justified by the fact that the Scenario II (Table \ref{tab_scen2}) is related to modeling structures having more parameters than those in Scenario I. It seems fair to say that this decrease is small and it does not compromise the inference in the general regression setting.

\section{Real Case Study: Analysis of IPASS Clinical Trial} 
\label{sec_applic}

In this section we present the analysis of the reconstructed IPASS clinical trial data reported in \cite{2015_Argyropoulos}. Although reconstructed, this data set preserves all features exhibited in references with full access to the observations from this clinical trial. The data base is related to the period of March 2006 to April 2008. The main purpose of the study is to compare the drug gefitinib against carboplatin/paclitaxel doublet chemotherapy as first line treatment, in terms of progression free survival (in months), to be applied to selected non-small-cell lung cancer (NSCLC) patients. According to the protocol established in the trial, $n = 1207$ previously untreated individuals in east Asia, who had advanced pulmonary adenocarcinoma and who were nonsmokers or former light smokers, were randomly assigned to receive either gefitinib (609 patients) or carboplatin + paclitaxel (608 patients). The observations indicate 965 occurrences of the event of interest ($79.3\%$), with 516 of them ($84.9\%$) corresponding to patients treated with carboplatin+paclitaxel, and 449 of them ($73.7\%$) being reported for those receiving gefitinib.  

The main aim here is to properly analyze the reconstructed IPASS data and estimate the crossing survival time using the models presented in Section \ref{sec_model}. In order to accomplish this goal, we consider both the likelihood-based and the Bayesian frameworks for the proposals $M_{1}$ and $M_{2}$. We emphasize that the same configurations adopted for the simulation study in Section \ref{sec_sim1} are applied in this real application; they include: polynomial's degree, bootstrap sample size, prior specifications and MCMC settings.

\begin{table}[h!]
\centering \footnotesize
\setlength\tabcolsep{2.8pt}
\caption{IPASS clinical trial analysis. Coefficients and the crossing survival time ($t^*$) summarized via point estimate (est.), standard error or standard deviation (se. and sd.), 95\% confidence interval (CI) or 95\% credibility interval with highest posterior density (HPD).} \vspace{5pt}
\begin{tabular}{lrrrrr}   \hline
  {\bf \small ML approach} & par  & est. & se. & lower$_{\text{\tiny CI}}$ & upper$_{\text{\tiny CI}}$ \\   \hline
&$\psi$  & 1.23 & 0.18 & 0.87 & 1.60 \\ 
$M_{1}$ & $\phi$  & -1.32 & 0.08 & -1.48 & -1.16 \\ 
& $t^{*}$ & 6.06  & 0.37 &  5.30 &  6.77 \\ \hline
& $\psi$  & 1.22 & 0.18 & 0.86 & 1.59 \\ 
$M_{2}$ & $\phi$  & -1.32 & 0.08 & -1.48 & -1.15 \\ 
& $t^{*}$ & 6.08  & 0.37 & 5.32  & 6.77 \\ \hline
  {\bf \small Bayesian} & par  & est. & sd. & lower$_{\text{\tiny HPD}}$ & upper$_{\text{\tiny HPD}}$ \\   \hline
&$\psi$  & 1.22 & 0.18  & 0.89 & 1.58 \\ 
$M_{1}$ & $\phi$  & -1.33 & 0.09 & -1.49 & -1.17 \\ 
& $t^{*}$ & 5.96 & 0.37 & 5.19 & 6.65 \\ \hline
& $\psi$  & 1.19 & 0.17 & 0.85 & 1.52 \\ 
$M_{2}$ & $\phi$  & -1.31 & 0.08 & -1.48 & -1.16 \\ 
& $t^{*}$ & 5.94 & 0.37 & 5.21 & 6.65 \\ \hline \\
\end{tabular}
\label{tab_ipass_coefs}
\end{table}

\begin{table}
\centering \footnotesize
\caption{IPASS clinical trial analysis. Short and long term hazard ratios summarized via point estimate (est.), standard error or standard deviation (se. and sd.), 95\% confidence interval (CI) or the 95\% credibility interval with highest posterior density (HPD).} \vspace{5pt}
\label{tab_ipass_hr}
\begin{tabular}{llrrrr}   \hline
 {\bf \small ML approach} & hazard ratio  & est. & se. & lower$_{\text{\tiny CI}}$ & upper$_{\text{\tiny CI}}$ \\  \hline
$M_{1}$ & gefitinib (short) & 3.47 & 0.64 & 2.40 & 4.96 \\ 
& carboplatin/paclitaxel (long) & 3.76 & 0.31 & 3.17 & 4.39 \\ \hline
$M_{2}$ & gefitinib (short)  & 3.44 & 0.64 & 2.36 & 4.90  \\ 
& carboplatin/paclitaxel (long) & 3.74 & 0.31 & 3.17 & 4.38 \\ \hline
 {\bf \small Bayesian} & hazard ratio  & est. & sd. & lower$_{\text{\tiny HPD}}$ & upper$_{\text{\tiny HPD}}$ \\  \hline
$M_{1}$ & gefitinib (short) & 3.43 & 0.62 & 2.28 & 4.62 \\ 
& carboplatin/paclitaxel (long) & 3.78 & 0.32 & 3.20 & 4.42 \\ \hline
$M_{2}$ & gefitinib (short)  & 3.34 & 0.58 & 2.27 & 4.51  \\ 
& carboplatin/paclitaxel (long) & 3.73 & 0.30 & 3.19 & 4.37 \\ \hline \\
\end{tabular}
\end{table}

\begin{figure}
\centering \footnotesize
$$
\begin{array}{cc}
(a) & (b) \vspace{-10pt} \\
\includegraphics[scale=0.35]{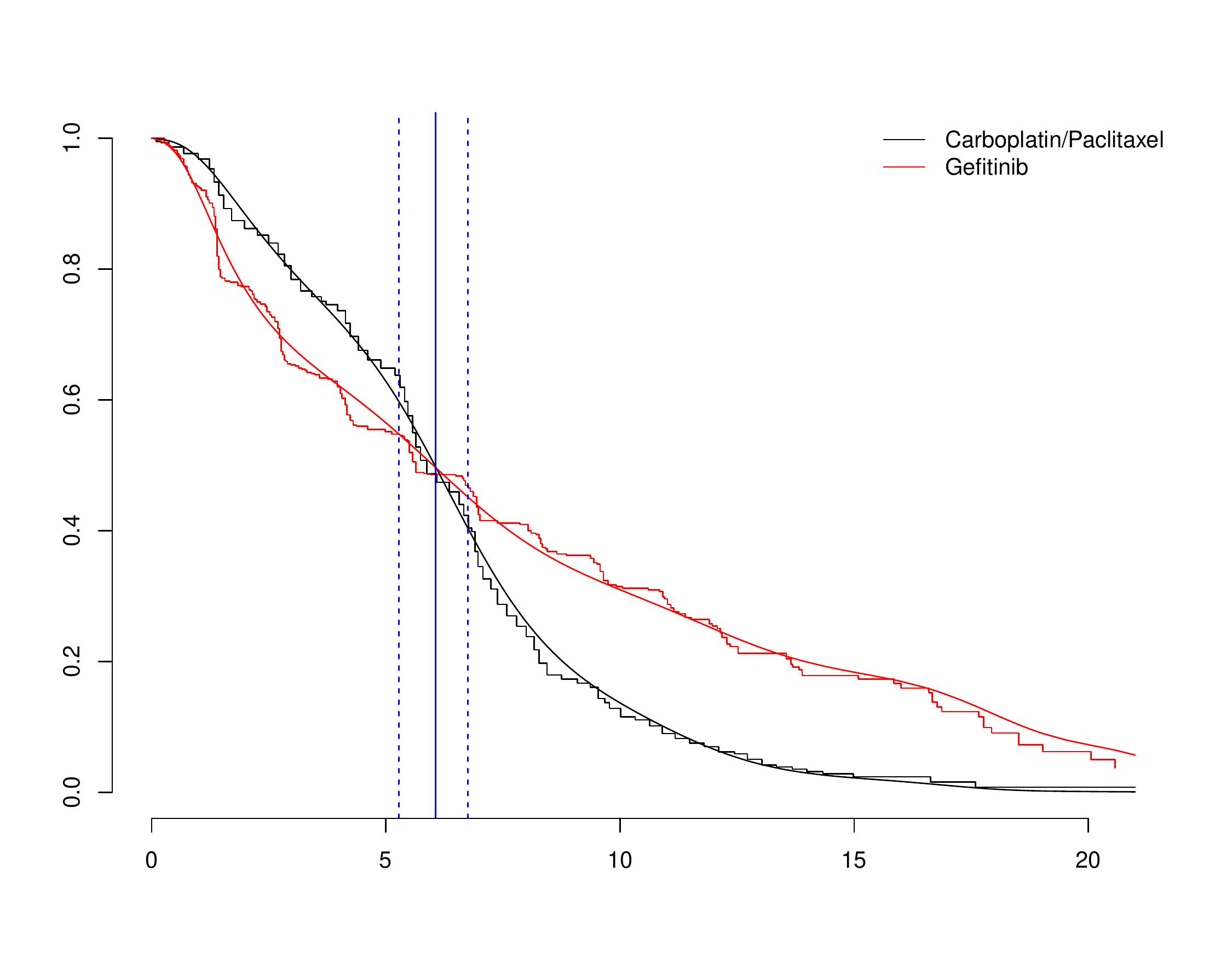} &
\includegraphics[scale=0.35]{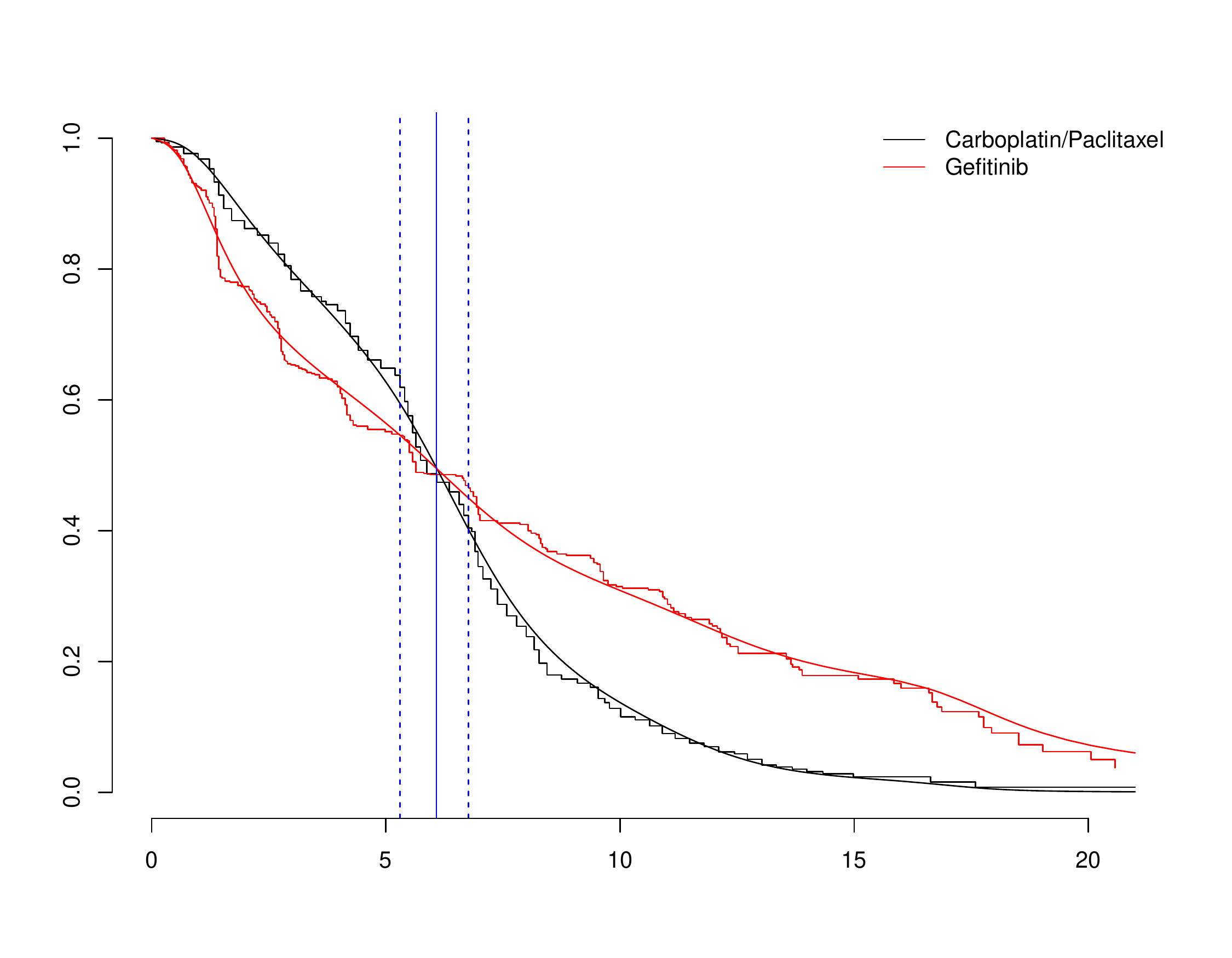} \\
(c) & (d)  \vspace{-10pt} \\
\includegraphics[scale=0.35]{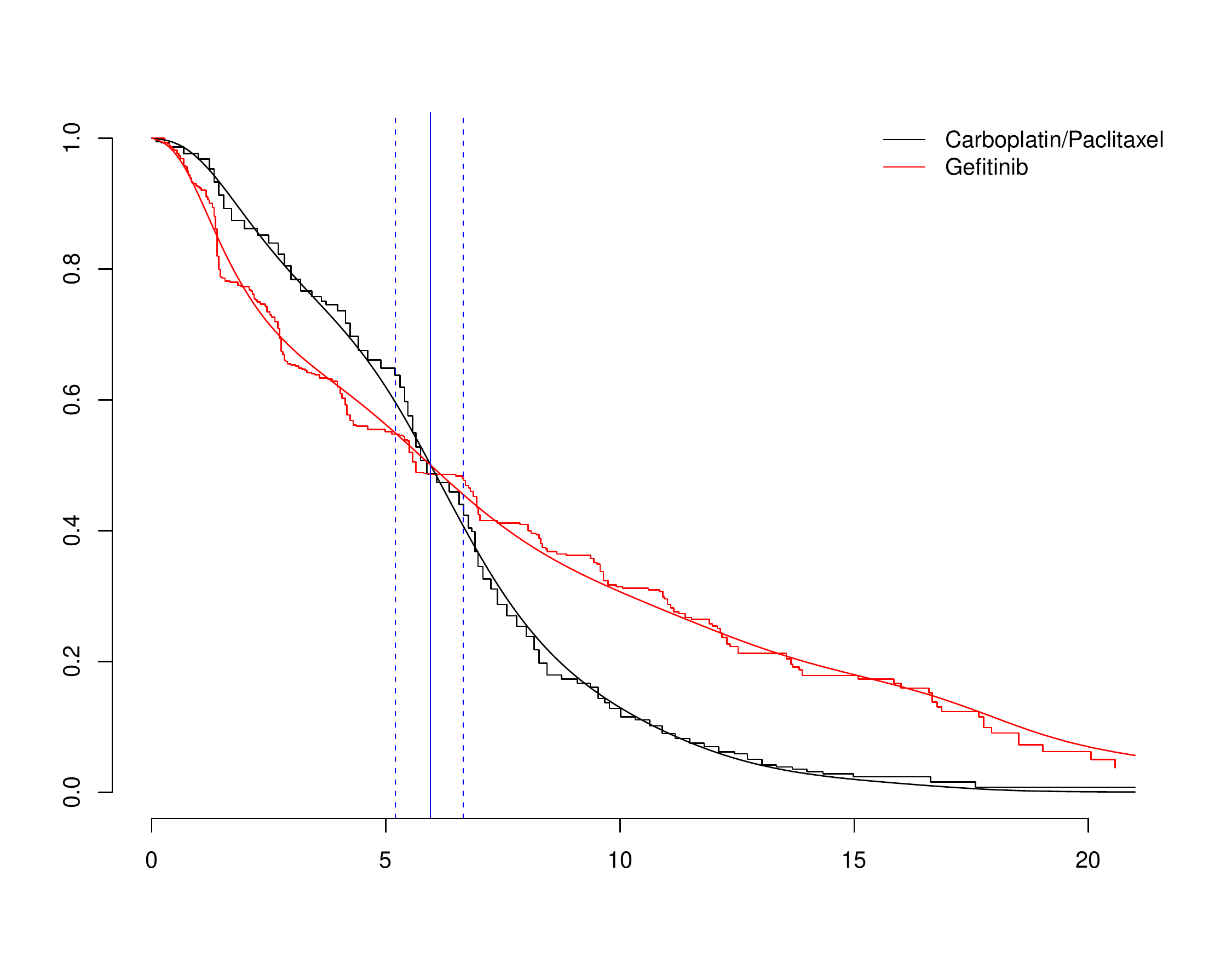} &
\includegraphics[scale=0.35]{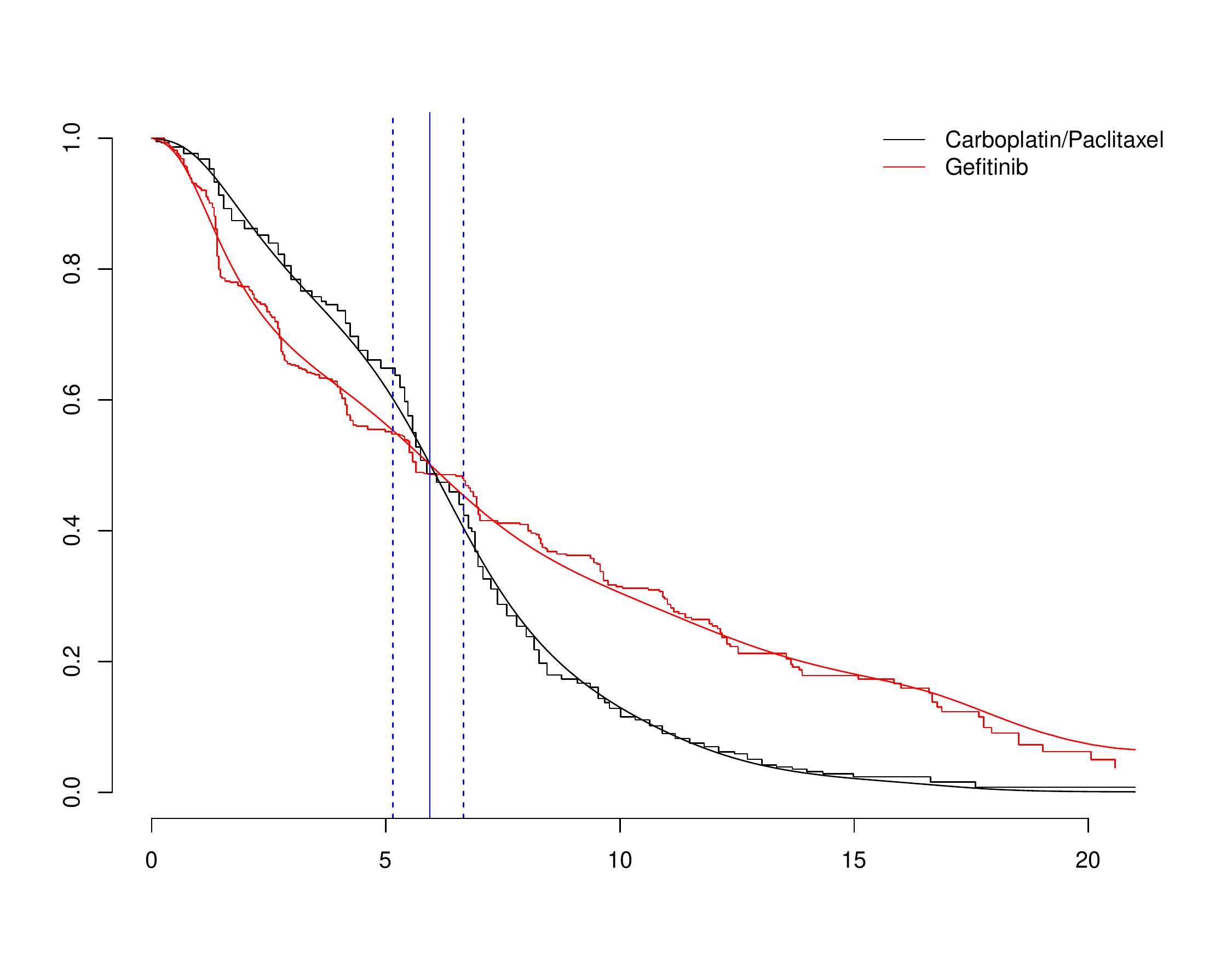} \\
\end{array}
$$
\vspace{-20pt}
\caption{Estimated survival curves - Kaplan-Meyer (step function) and proposed models (continuous lines). Consider $M_1$ via ML in panel ($a$), Bayesian $M_1$ in ($b$), $M_2$ via ML in ($c$) and Bayesian $M_2$ in ($d$). The horizontal axis represents the follow up time in months. The estimated crossing time is given by the vertical continuous line (dashed lines indicates the 95\% confidence or credibility interval).}
\label{fig_ht}
\end{figure}

Table \ref{tab_ipass_coefs} provides the estimates for the short/long-term regression coefficients and the crossing survival time, along with their respective standard errors/deviations and corresponding interval estimates obtained by the four fitted models. As expected, the short/long-term regression coefficient estimates have opposite signs, implying that the survival curves associated with gefitinib and carboplatin/paclitaxel treatments must cross each other at some point. In addition, both $\psi$ and $\phi$ are statistically significant, indicating that there is a non-negligible difference between these treatments when comparing the early and late follow up periods. This is confirmed through the analysis of Table \ref{tab_ipass_hr}, which shows estimated hazard ratios. Note that the hazard of a patient receiving gefitinib at the begining of the treatment is more than three times the hazard of those treated with carboplatin/paclitaxel. On the other hand, gefitinib tends to improve in the long run with respect to carboplatin+paclitaxel; the hazard ratios between patients treated with carboplatin/paclitaxel and gefitinib are larger than 3.5.

The results displayed in Tables \ref{tab_ipass_coefs} and \ref{tab_ipass_hr} clearly shows the serious implications of the misleading use of the Cox PH model when dealing with lifetime data with crossing survival curves. Our study indicates that the correct clinical practice to be adopted, in terms of choosing gefitinib or carboplatin/paclitaxel, should take into account the inversion of the effectiveness of these treatments. Specifically, according to Table \ref{tab_ipass_coefs}, patients should start their treatment receiving carboplatin+paclitaxel, and then switch to gefitinib after 6 months of follow up.

Figure \ref{fig_ht} compares the survival curves obtained through the proposed models and the usual Kaplan-Meyer estimator for each treatment group. Note that, for all cases, the trajectory of the estimated continuous curves are quite similar to the one exhibited by the step function. The crossing survival time $t^*$ is estimated near 6 months (see vertical lines) with 95\% interval reasonably small suggesting some precision. The behavior of the curves clearly indicates that the two treatments experience an inversion along the time, with gefitinib associated to the lower survival in the short-term or initial phase of the study.

\section{Conclusions} \label{sec_conclusions}

This paper introduced an unified approach to fit the YP model by modeling either the baseline hazard function or the baseline odds function via Bernstein polynomials, under both the likelihood-based and the Bayesian frameworks. The use of BP to manage the baseline functions yields some flexibility as no constrains on the shape of these functions are required. The existence of closed form expressions for the baseline functions determines simple formulations for the likelihood which, in turn, considerably simplifies all the inferential procedure. Another central characteristic of the proposed models is their ability to accurately estimate crossing survival times in a straightforward way. We believe that this feature is extremely important in many real applications where detecting the crossing survival time is a key information to establish guidelines for a treatment. This is well illustrated here through the analysis of the reconstructed IPASS data. In addition, this work also provides an alternative formulation for the YP model allowing the effects of some covariates to remain constant over time. This possibility is only mentioned in \cite{2005_YangPrentice} and, as far as we know, it has not been addressed in the literature until now.

We evaluated the performance of the proposed models through a comprehensive Monte Carlo simulation study considering a two-sample situation, and a general regression scenario with a set of covariates. To the best of our knowledge, such an extensive simulation study involving the YP model has never been conducted in the literature. Overall, the proposed models presented good performance in terms of relative bias and coverage probabilities. The real illustration related to the IPASS data set revealed that it is a mistake to trust the analysis based on the usual PH model. This choice induces wrong conclusions which, in turn, may lead to inappropriate clinical practices in terms of the best treatment for the patients. As demonstrated in our analysis, in order to increase progression-free survival probabilities, patients should start their treatment receiving carboplatin/paclitaxel, and then switch to gefitinib after approximately 6 months of treatment.

Future research includes the development of residual analysis techniques and diagnostic measures to assess the goodness of fit of the proposed models. In upcoming works we intend to extend the proposed models to account for interval-censored data and survival data with cure fraction.


\section*{Acknowledgements}
The second author gratefully acknowledge the support from Fundação de
Amparo a Pesquisa do Estado de Minas Gerais (FAPEMIG).




\bibliographystyle{biom} \bibliography{references}

\begin{thebibliography}{}

\bibitem[\protect\citeauthoryear{Argyropoulos and Unruh}{Argyropoulos and
  Unruh}{2015}]{2015_Argyropoulos}
Argyropoulos, C. and Unruh, M.~L. (2015).
\newblock Analysis of time to event outcomes in randomized controlled trials by
  generalized additive models.
\newblock {\em PLOS One} {\bf 10,} 1--33.

\bibitem[\protect\citeauthoryear{Babu, Canty, and Chaubey}{Babu
  et~al.}{2002}]{Babu02}
Babu, G., Canty, A., and Chaubey, Y. (2002).
\newblock Application of {Bernstein} polynomials for smooth estimation of
  distribution and density function.
\newblock {\em Journal of Statistical Planning and Inference} {\bf 105,}
  377--392.

\bibitem[\protect\citeauthoryear{Bennett}{Bennett}{1983}]{Bennett83}
Bennett, S. (1983).
\newblock Analysis of survival data by the proportional odds model.
\newblock {\em Statistics in Medicine} {\bf 2,} 273--277.

\bibitem[\protect\citeauthoryear{Breslow, Edler, and Berger}{Breslow
  et~al.}{1984}]{1984_Breslow}
Breslow, N.~E., Edler, L., and Berger, J. (1984).
\newblock A two-sample censored-data rank test for acceleration.
\newblock {\em Biometrics} {\bf 40,} 1049--1062.

\bibitem[\protect\citeauthoryear{Chang, Chien, Hsiung, Wen, and Wu}{Chang
  et~al.}{2007}]{Chang07}
Chang, I.~S., Chien, L.~C., Hsiung, C.~A., Wen, C.~C., and Wu, Y.~J. (2007).
\newblock {\em Complex datasets and inverse problems}, volume~54, chapter Shape
  restricted regression with random {Bernstein} polynomials, pages 187--202.
\newblock Institute of Mathematical Statistics, Hayward.

\bibitem[\protect\citeauthoryear{Chang, Hsiung, Wu, and Yang}{Chang
  et~al.}{2005}]{Chang05}
Chang, I.~S., Hsiung, C.~A., Wu, Y.~J., and Yang, C.~C. (2005).
\newblock Bayesian survival analysis using {B}ernstein polynomials.
\newblock {\em Scandinavian Journal of Statistics} {\bf 32,} 447--466.

\bibitem[\protect\citeauthoryear{Chen, Hanson, and Zhang}{Chen
  et~al.}{2014}]{Chen14}
Chen, Y., Hanson, T., and Zhang, J. (2014).
\newblock Accelerated hazard model based on parametric families generalized
  with {B}ernstein polynomials.
\newblock {\em Biometrics} {\bf 70,} 192--201.

\bibitem[\protect\citeauthoryear{Choudhuri, Ghosal, and Roy}{Choudhuri
  et~al.}{2004}]{Choud04}
Choudhuri, N., Ghosal, S., and Roy, A. (2004).
\newblock Bayesian estimation of the spectral density of a time series.
\newblock {\em Journal of the American Statistical Association} {\bf 99,}
  1050--1059.

\bibitem[\protect\citeauthoryear{Cox}{Cox}{1972}]{Cox1972}
Cox, D.~R. (1972).
\newblock Regression models and life-tables.
\newblock {\em Journal of the Royal Statistical Society, Series B} {\bf 34,}
  187--220.
\newblock with discussion.

\bibitem[\protect\citeauthoryear{Diao, Zeng, , and Yang}{Diao
  et~al.}{2013}]{2013_Diao}
Diao, G., Zeng, D., , and Yang, S. (2013).
\newblock Efficient semiparametric estimation of short-term and long-term
  hazard ratios with right-censored data.
\newblock {\em Biometrics} {\bf 69,} 840--849.

\bibitem[\protect\citeauthoryear{Homan and Gelman}{Homan and
  Gelman}{2014}]{2014_NUTS}
Homan, M.~D. and Gelman, A. (2014).
\newblock {The No-U-Turn Sampler: Adaptively Setting Path Lengths in
  Hamiltonian Monte Carlo}.
\newblock {\em Journal of Machine Learning Research} {\bf 15,} 1351--1381.

\bibitem[\protect\citeauthoryear{Kalbfleisch and Prentice}{Kalbfleisch and
  Prentice}{2002}]{Kalb2002}
Kalbfleisch, J.~D. and Prentice, R.~L. (2002).
\newblock {\em The Statistical Analysis of Failure Time Data}.
\newblock Wiley Series in Probability and Statistics. Wiley, New York.

\bibitem[\protect\citeauthoryear{Lorentz}{Lorentz}{2012}]{Lorentz12}
Lorentz, G.~G. (2012).
\newblock {\em Bernstein polynomials}.
\newblock AMS Chelsea Publishing. American Mathematical Society, New York, 2
  edition.

\bibitem[\protect\citeauthoryear{Mok, Wu, Thongprasert, Yang, Chu, Saijo,
  Sunpaweravong, Han, Margono, Ichinose, Nishiwaki, Ohe, Yang, Chewaskulyong,
  Jiang, Duffield, Watkins, Armour, and Fukuoka}{Mok et~al.}{2009}]{2009_Mok}
Mok, T.~S., Wu, Y.~L., Thongprasert, S., Yang, C.~H., Chu, D.~T., Saijo, N.,
  Sunpaweravong, P., Han, B., Margono, B., Ichinose, Y., Nishiwaki, Y., Ohe,
  Y., Yang, J.~J., Chewaskulyong, B., Jiang, H., Duffield, E.~L., Watkins,
  C.~L., Armour, A.~A., and Fukuoka, M. (2009).
\newblock Gefitinib or carboplatin-paclitaxel in pulmonary adenocarcinoma.
\newblock {\em New England Journal of Medicine} {\bf 361,} 947--957.
\newblock PMID: 19692680.

\bibitem[\protect\citeauthoryear{Nieto-Barajas}{Nieto-Barajas}{2014}]{2014_Barajas}
Nieto-Barajas, L.~E. (2014).
\newblock Bayesian semiparametric analysis of short- and long-term hazard
  ratios with covariates.
\newblock {\em Computational Statistics and Data Analysis} {\bf 71,} 477--490.

\bibitem[\protect\citeauthoryear{Osman and Ghosh}{Osman and
  Ghosh}{2012}]{2012_OsmanGhosh}
Osman, M. and Ghosh, S.~K. (2012).
\newblock Nonparametric regression models for right-censored data using
  {B}ernstein polynomials.
\newblock {\em Computational Statistics and Data Analysis} {\bf 56,} 559--573.

\bibitem[\protect\citeauthoryear{Petrone}{Petrone}{1999}]{Petrone99}
Petrone, S. (1999).
\newblock Bayesian density estimation using {Bernstein} polynomials.
\newblock {\em Canadian Journal of Statistics} {\bf 27,} 105--126.

\bibitem[\protect\citeauthoryear{{R Core Team}}{{R Core Team}}{2019}]{cran}
{R Core Team} (2019).
\newblock {\em R: A Language and Environment for Statistical Computing}.
\newblock R Foundation for Statistical Computing, Vienna, Austria.

\bibitem[\protect\citeauthoryear{{Stan Development Team}}{{Stan Development
  Team}}{2018a}]{rstan}
{Stan Development Team} (2018a).
\newblock {RStan}: the {R} interface to {Stan}.
\newblock R package version 2.18.2.

\bibitem[\protect\citeauthoryear{{Stan Development Team}}{{Stan Development
  Team}}{2018b}]{Stan}
{Stan Development Team} (2018b).
\newblock {\em Stan Modeling Language Users Guide and Reference Manual},
  version 2.18.0 edition.

\bibitem[\protect\citeauthoryear{Tenbusch}{Tenbusch}{1997}]{Tenbusch97}
Tenbusch, A. (1997).
\newblock Nonparametric curve estimation with {Bernstein} estimates.
\newblock {\em Metrika} {\bf 45,} 1--30.

\bibitem[\protect\citeauthoryear{Tong, Zhu, and Sun}{Tong
  et~al.}{2007}]{2007_Tong}
Tong, X., Zhu, C., and Sun, J. (2007).
\newblock Semiparametric regression analysis of two-sample current status data,
  with applications to tumorigenicity experiments.
\newblock {\em Canadian Journal of Statistics} {\bf 35,} 575--584.

\bibitem[\protect\citeauthoryear{Vitale}{Vitale}{1975}]{Vitale75}
Vitale, R.~A. (1975).
\newblock A {B}ernstein polynomial approach to density function estimation.
\newblock {\em Stochastic Processes and Related Topics} {\bf 2,} 87--100.

\bibitem[\protect\citeauthoryear{Yang, , and Zhao}{Yang
  et~al.}{2012}]{2012_Yang}
Yang, S., , and Zhao, Y. (2012).
\newblock Checking the short-term and long-term hazard ratio model for survival
  data.
\newblock {\em Scandinavian Journal of Statistics} {\bf 39,} 554--567.

\bibitem[\protect\citeauthoryear{Yang and Prentice}{Yang and
  Prentice}{2005}]{2005_YangPrentice}
Yang, S. and Prentice, R.~L. (2005).
\newblock Semiparametric analysis of short-term and long-term hazard ratios
  with two-sample survival data.
\newblock {\em Biometrika} {\bf 92,} 1--17.

\bibitem[\protect\citeauthoryear{Yang and Prentice}{Yang and
  Prentice}{2011}]{2011_Yang}
Yang, S. and Prentice, R.~L. (2011).
\newblock Estimation of the 2-sample hazard ratio function using a
  semiparametric model.
\newblock {\em Biostatistics} {\bf 12,} 354--368.

\bibitem[\protect\citeauthoryear{Zhang, Wang, and Sun}{Zhang
  et~al.}{2017}]{2017_Zhang}
Zhang, H., Wang, P., and Sun, J. (2017).
\newblock Regression analysis of interval-censored failure time data with
  possibly crossing hazards.
\newblock {\em Statistics in Medicine} {\bf 37,} 768--775.

\bibitem[\protect\citeauthoryear{Zhou and Hanson}{Zhou and
  Hanson}{2017}]{Zhou17}
Zhou, H. and Hanson, T. (2017).
\newblock A unified framework for fitting {B}ayesian semiparametric models to
  arbitrarily censored survival data, including spatially-referenced data.
\newblock {\em Journal of the American Statistical Association} {\bf 113,}
  571--581.

\bibitem[\protect\citeauthoryear{Zhou, Hu, and Sun}{Zhou
  et~al.}{2017}]{QZhou17}
Zhou, Q., Hu, T., and Sun, J. (2017).
\newblock A {Sieve} semiparametric maximum likelihood approach for regression
  analysis of bivariate interval-censored failure time data.
\newblock {\em Journal of the American Statistical Association} {\bf 112,}
  664--672.

\end{thebibliography}



\label{lastpage}

\end{document}